\documentclass[letterpaper, 12pt, onecolumn,conference]{IEEEtran}
\IEEEoverridecommandlockouts
\usepackage{times}
\usepackage{epstopdf}
\usepackage{color}
\usepackage{fix2col}
\usepackage{mathdots}
\usepackage[T1]{fontenc}

\usepackage{rotate}

\usepackage[english]{babel}
\usepackage{cite}      %
\usepackage[pdftex]{graphicx}
\usepackage{url}       %
\usepackage{amssymb}
\usepackage{amsmath}
\usepackage{array}     %
\usepackage{times}
\usepackage{footnote}  
\usepackage{xspace}
\usepackage{setspace}
\usepackage[small,normal,bf,up]{caption2}

\usepackage[total={6in,9in},top=1in,left=1.25in]{geometry}

\renewcommand{\sf}{\fontsize{6pt}{6pt}\selectfont}

\newtheorem{definition}{Definition}[section]
\newtheorem{theorem}{Theorem}[section]

\newtheorem{lemma}{Lemma}[section]

\title{\huge Distributed Remote Vector Gaussian Source Coding for Wireless Acoustic Sensor Networks\\[5mm]}

\author{\IEEEauthorblockN{Adel Zahedi\IEEEauthorrefmark{1},
Jan \O stergaard\IEEEauthorrefmark{1},
S\o ren Holdt Jensen\IEEEauthorrefmark{1}, 
Patrick Naylor\IEEEauthorrefmark{2}, \\
and S\o ren Bech\IEEEauthorrefmark{1}\IEEEauthorrefmark{3}}
\thanks{The research leading to these results has received funding from the European Union's Seventh Framework Programme (FP7/2007-2013) under grant agreement n$^\circ$ ITN-GA-2012-316969.}
\IEEEauthorblockA{\IEEEauthorrefmark{1}Department of Electronic Systems\\
Aalborg University, 9220 Aalborg, Denmark \\
Email: \{adz, jo, shj, sbe\}@es.aau.dk}
\IEEEauthorblockA{\IEEEauthorrefmark{2}Electrical and Electronic Engineering Department\\
London Imperial College, London SW7 2AZ, UK\\
Email: p.naylor@imperial.ac.uk}
\IEEEauthorblockA{\IEEEauthorrefmark{3}Bang \& Olufsen\\
7600 Struer, Denmark}}

\begin{document}

\maketitle

\vspace{2.5cm}

\begin{abstract}
In this paper, we consider the problem of remote vector Gaussian source coding for a wireless acoustic sensor network. Each node receives messages from multiple nodes in the network and decodes these messages using its own measurement of the sound field as side information. The node's measurement and the estimates of the source resulting from decoding the received messages are then jointly encoded and transmitted to a neighboring node in the network. We show that for this distributed source coding scenario, one can encode a so-called conditional sufficient statistic of the sources instead of jointly encoding multiple sources. We focus on the case where node measurements are in form of noisy linearly mixed combinations of the sources and the acoustic channel mixing matrices are invertible. For this problem, we derive the rate-distortion function for vector Gaussian sources and under covariance distortion constraints.
\end{abstract}

\newpage


\section{Introduction}

A Wireless Acoustic Sensor Network (WASN) is a set of wireless microphones equipped with some communication, signal processing, and possibly memory units, which are randomly scattered in the environment. The communication unit allows a sensor node to communicate with a base station and also with other nodes, and the signal processing unit enables a node to perform local processing. The random structure of the network removes the array-size limitations imposed by classical microphone arrays, thereby providing better performance in sense of high-SNR spatio-temporal sampling of the sound field and spatial diversity. The interested reader is referred to \cite{Bertrand} for a review of WASNs.

\begin{figure}[tb]
\begin{center}
\includegraphics[scale=0.28]{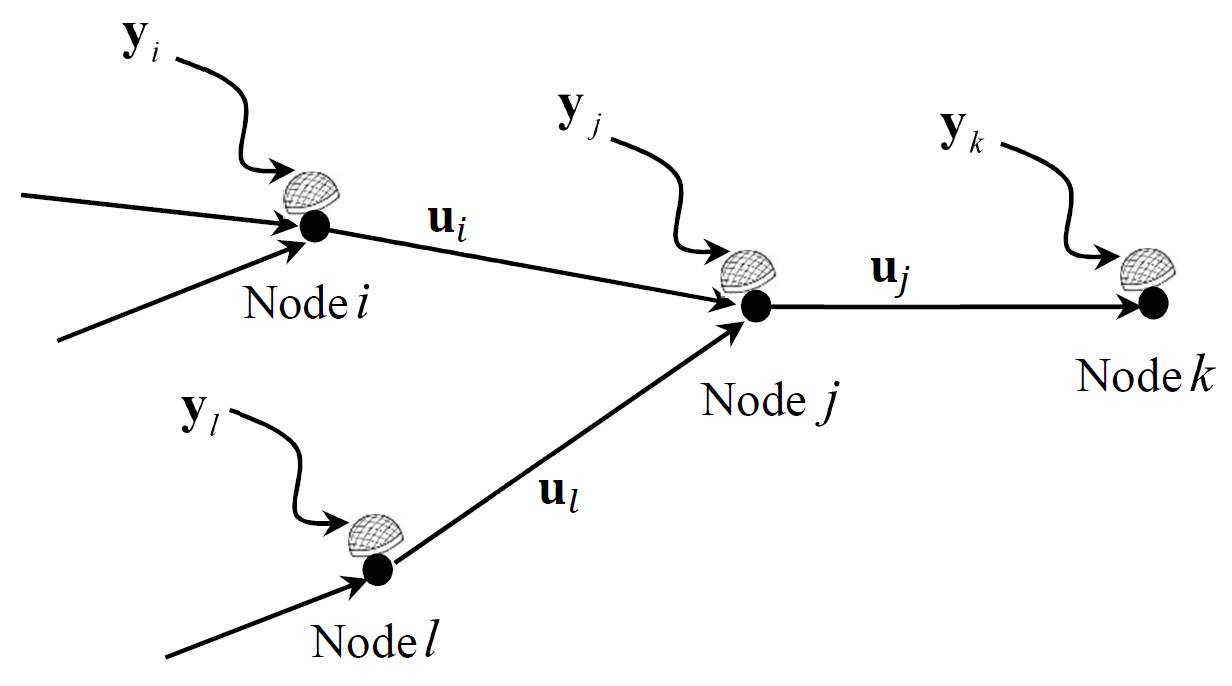}
\caption{A schematic graph of the WASN}
\label{net}
\end{center}
\end{figure}

Due to strict power and bandwidth constraints on wireless microphones, it might not be possible for nodes which are far from the base station to directly deliver their messages. A multi-hop scenario where the message goes along adjacent nodes until reaching the base-station is a commonplace alternative. We assume that a node is supposed to receive messages from multiple neighboring nodes and then combine them with its own measurement of the sound field into a new message to be forwarded through the network. This is illustrated in Fig.\ref{net}.

In \cite{Jan}, this problem was solved for scalar Gaussian sources in a non-distributed setting i.e., without making use of the availability of the nodes' measurements as side information. It was shown that coding at intermediate nodes can result in significant gains in terms of sum-rate or distortion. In this paper, we consider vector Gaussian sources and also take into account the fact that messages received by a node are correlated with the node's measurement of the sound field. Thus, we consider a distributed scenario, and make use of the destination node's measurement as side information to decrease the required rate for transmission to each node \cite{SW}, \cite{Wyner1},\cite{Wyner2}.


We derive the rate-distortion (RD) function for an arbitrary node in the network with a distortion constraint defined in form of a covariance matrix, cf. \cite{Rahman}, \cite{Zhang}. In Section II, we introduce the notation and formulate the problem, which turns out to involve joint coding of multiple sources. In Section III, we derive a conditional sufficient statistic for multiple Gaussian sources and show that for the above-mentioned distributed source coding (DSC) problem, one can encode a conditional sufficient statistic instead of joint encoding of multiple sources. The RD function for the resulting problem will be derived in Section IV. The paper is concluded in Section V.

\section{Notation and Problem Formulation}

\begin{figure}[th]
\begin{center}
\includegraphics[scale=0.33]{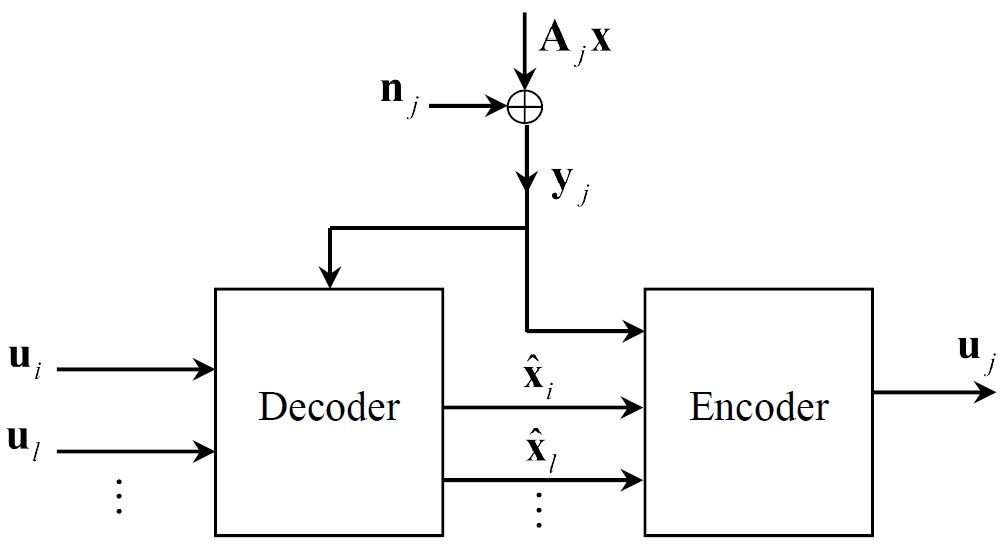}
\caption{Node $j$ of the network}
\label{1node}
\end{center}
\end{figure}

We denote by $I(\cdot;\cdot), h(\cdot),$ and $E[\cdot]$ the information theoretic operations of mutual information, differential entropy, and expectation, respectively. Probability density functions are denoted by $f(\cdot)$ and covariance and cross-covariance matrices are denoted by $\bf{\Sigma}$. We assume that all covariance matrices are of full rank. We denote Markov chains by two-headed arrows; e.g. ${\bf{u}} \leftrightarrow {\bf{v}} \leftrightarrow {\bf{w}}$. We assume that the source generates independent Gaussian vectors of length $n$ denoted by $\bf{x} \in \mathbb{R}^\mathnormal{n}$. While the vectors are independent, the components of each vector are correlated. Node $j$ makes a noisy measurement ${\bf{y}}_j$ of the source given by:

\begin{equation}
\label{measure}
{{\bf{y}}_j} = {{\bf{A}}_j}{\bf{x}} + {{\bf{n}}_j},
\end{equation}

\noindent
where ${\bf{n}}_j$ is the additive Gaussian noise at node $j$ and the matrix ${\bf{A}}_j$ models the mixing effect of the acoustic channel. The noise is assumed to be independent of $\bf{x}$. Node $j$ also receives messages from nodes $i,l,...$ denoted by ${\bf{u}}_i,{\bf{u}}_l,...$, from which it can make estimations ${{{\bf{\hat x}}}_i},{{{\bf{\hat x}}}_l},...$ of the source by decoding the messages with ${\bf{y}}_j$ as the decoder side information. The problem is to find the minimum rate $R_j$ to jointly encode ${\bf{y}}_j,{{{\bf{\hat x}}}_i},{{{\bf{\hat x}}}_l},...$ into a message ${\bf{u}}_j$ to be sent to node $k$, while satisfying the given distortion constraint ${\bf{D}}_j$ and considering ${\bf{y}}_k$ as side information. This is illustrated in Figs.\ref{net} and \ref{1node}.

Note that one could further decrease the network sum-rate by taking into account the correlation between the messages which are sent to a common node. However, we leave out this possibility in this paper.

Throughout this work, we assume that the acoustic mixing matrices are fixed and known. We also assume that joint statistics of the source and the noise variances are available. Finally, although the model in (\ref{measure}) is appropriate for acoustic networks, we do not consider real acoustic signals in this work. Instead, we consider Gaussian sources for simplicity of mathematical analysis. The case of real audio measurements is the focus of our future work.


\section{Conditional Sufficient Statistics}

Assume that $\bf{y}$ is a random vector or a collection of random vectors with probability density function $f(\bf{y})$.

\begin{definition}\label{define}
A sufficient statistic for the estimation of $\bf{x}$ from $\bf{y}$ is a function $T(\bf{y})$ of $\bf{y}$ for which $f({\bf{y}}|T(\bf{y}),\bf{x})$ is not a function of $\bf{x}$. In other words, ${\bf{y}} \leftrightarrow {T(\bf{y})} \leftrightarrow {\bf{x}}$.
\end{definition}

\begin{theorem}[Neyman-Fisher Factorization Theorem \cite{Kay}]\label{suff}
$T(\bf{y})$ is a sufficient statistic of $\bf{y}$ for estimating $\bf{x}$ if and only if $f(\bf{y}|\bf{x})$ can be factorized as:

\begin{equation}
\label{factor}
f\left( {{\bf{y}}|{\bf{x}}} \right) = p\left( {\bf{y}} \right)q\left( {T\left( {\bf{y}} \right),{\bf{x}}} \right),
\end{equation}

\noindent
where $p$ (depending on $\bf{y}$ only) and $q$ (depending on $\bf{x}$ and $T(\bf{y})$ but not on $\bf{y}$ directly) are nonnegative functions.

\end{theorem}

The Factorization Theorem enables us to find a sufficient statistic for e.g. random vectors in Gaussian noise as shown in the following lemma\footnote{To the best of our knowledge, no proof for this lemma appears in the literature. We provide a proof here for completeness.}.

\begin{lemma}\label{Gauss}
If ${\bf{y}}_j; \: j = 1,2,...,N$ are measurements of a random vector $\bf{x}$ in mutually independent Gaussian noises as in (\ref{measure}), then $T\left( {{{\bf{y}}_1},...,{{\bf{y}}_N}} \right)$ is a sufficient statistic of ${\bf{y}}_j$ for estimating $\bf{x}$, where

\begin{equation}
\label{Gauss_suff}
T\left( {{{\bf{y}}_1},...,{{\bf{y}}_N}} \right) = \sum\limits_{j = 1}^N {{\bf{A}}_j^T{\bf{\Sigma }}_j^{ - 1}{{\bf{y}}_j}},
\end{equation}

\noindent
and ${\bf{\Sigma }}_j$ is the covariance matrix of the noise ${\bf{n}}_j$.

\end{lemma} 

\begin{IEEEproof}
We prove the lemma for $N=2$. The proof for the general case is similar. 

Since ${\bf{n}}_1$ and ${\bf{n}}_2$ are independent, the joint conditional density function of ${\bf{y}}_1$ and ${\bf{y}}_2$ can be written as:

\begin{equation}
\label{joint}
f\left( {{{\bf{y}}_1},{{\bf{y}}_2}|{\bf{x}}} \right) = \alpha \exp \left( { - \frac{1}{2}\varphi } \right),
\end{equation}

\noindent
where $\alpha$ is independent of $\bf{x}$, ${\bf{y}}_1$ and ${\bf{y}}_2$, and $\varphi$ is defined as:

\begin{equation}
\label{phi}
\varphi  = {\left( {{{\bf{y}}_1} - {\bf{x}}} \right)^T}{\bf{\Sigma }}_1^{ - 1}\left( {{{\bf{y}}_1} - {\bf{x}}} \right) + {\left( {{{\bf{y}}_2} - {\bf{x}}} \right)^T}{\bf{\Sigma }}_2^{ - 1}\left( {{{\bf{y}}_2} - {\bf{x}}} \right).
\end{equation}

\noindent
Expanding $\varphi$, rearranging the terms, and substituting in (\ref{joint}), leads to

\begin{eqnarray}
f\left( {{{\bf{y}}_1},{{\bf{y}}_2}|{\bf{x}}} \right) = \alpha \exp \left( { - \frac{{{\bf{y}}_1^T{\bf{\Sigma }}_1^{ - 1}{{\bf{y}}_1} + {\bf{y}}_2^T{\bf{\Sigma }}_2^{ - 1}{{\bf{y}}_2}}}{2}} \right) \nonumber \\
\times \exp \left( { - \frac{{{{\bf{x}}^T}\left( {{\bf{\Sigma }}_1^{ - 1} + {\bf{\Sigma }}_2^{ - 1}} \right){\bf{x}}}}{2}} \right)\exp \left( {\frac{{{T^T}{\bf{x}} + {{\bf{x}}^T}T}}{2}} \right),
\end{eqnarray}

\noindent
which shows that $T$ is a sufficient statistic according to the Factorization Theorem.

\end{IEEEproof}

To make the notion of sufficient statistics applicable to our DSC problem, we need to introduce a conditional version:

\begin{definition}\label{define1}
A conditional sufficient statistic for the estimation of $\bf{x}$ from $\bf{y}$ given $\bf{z}$ is a function $T(\bf{y})$ of $\bf{y}$ for which $f(\bf{y}|T(\bf{y}),\bf{x},\bf{z})$ is not a function of $\bf{x}$. 
\end{definition}

\begin{lemma}\label{Gauss1}
If ${\bf{y}} \leftrightarrow {\bf{x}} \leftrightarrow {\bf{z}}$, then any sufficient statistic $T(\bf{y})$ of $\bf{y}$ for estimating $\bf{x}$ is also a conditional sufficient statistic of $\bf{y}$ for estimating $\bf{x}$ given $\bf{z}$.
\end{lemma}

\begin{IEEEproof}

\begin{eqnarray}
f\left( {{\bf{y}}|T\left( {\bf{y}} \right),{\bf{x}},{\bf{z}}} \right) &&= \frac{{f\left( {{\bf{y}}|{\bf{x}},{\bf{z}}} \right)}}{{f\left( {T\left( {\bf{y}} \right)|{\bf{x}},{\bf{z}}} \right)}} \nonumber \\
 &&= \frac{{f\left( {{\bf{y}}|{\bf{x}}} \right)}}{{f\left( {T\left( {\bf{y}} \right)|{\bf{x}}} \right)}} \nonumber \\
 &&= f\left( {{\bf{y}}|T\left( {\bf{y}} \right),{\bf{x}}} \right) \nonumber \\
 &&= f\left( {{\bf{y}}|T\left( {\bf{y}} \right)} \right)
\end{eqnarray}

\noindent
which is independent of $\bf{x}$.

\end{IEEEproof}

\begin{lemma}\label{lem3}
Given side information $\bf{z}$ at the decoder, there is no loss in terms of rate and distortion by encoding a conditional sufficient statistic of ${{{\bf{y}}_1},...,{{\bf{y}}_N}}$ given $\bf{z}$ instead of joint encoding of these sources.
\end{lemma} 

\begin{IEEEproof}
The proof follows along the lines of the proof for the unconditional case presented in \cite{Eswaran}, and is therefore omitted.

\end{IEEEproof}

Combining the results of this section, we have the following theorem for the problem formulated in Section II:

\begin{theorem}\label{th1}
Given side information ${\bf{y}}_k$ at the decoder, the RD function for the problem of joint encoding of multiple sources ${\bf{y}}_j,{{{\bf{\hat x}}}_i},{{{\bf{\hat x}}}_l},...$ coincides with the RD function for the DSC problem of encoding the single source:

\begin{equation}
\label{T}
{T_j} = {\bf{A}}_j^T{\bf{\Sigma }}_j^{ - 1}{{\bf{y}}_j} + {\bf{D}}_i^{ - 1}{{{\bf{\hat x}}}_i} + {\bf{D}}_l^{ - 1}{{{\bf{\hat x}}}_l} + ... 
\end{equation}

\noindent
where ${\bf{D}}_i$ is the distortion matrix for node $i$.

\end{theorem}

\begin{IEEEproof}
Using the backward channel model we have ${\bf{x}} = {{{\bf{\hat x}}}_i} + {{\bf{q}}_i}$ where the covariance matrix of ${{\bf{q}}_i}$ is ${\bf{D}}_i$. This means that ${{{\bf{\hat x}}}_i}$ can be written as:

\begin{equation}
\label{xhat}
{{{\bf{\hat x}}}_i} = {{\bf{H}}_i}{\bf{x}} + {{\bf{\eta }}_i},
\end{equation}

\noindent
where

\begin{eqnarray}
\label{xhat1}
&&{{\bf{H}}_i} = \left( {{{\bf{\Sigma }}_x} - {{\bf{D}}_i}} \right){\bf{\Sigma }}_x^{ - 1}, \\
\label{xhat2}
&&{{\bf{\Sigma }}_{{{\bf{\eta }}_i}}} = \left( {{{\bf{\Sigma }}_x} - {{\bf{D}}_i}} \right){\bf{\Sigma }}_x^{ - 1}{{\bf{D}}_i}.
\end{eqnarray}

\noindent
Substituting (\ref{xhat1}) and (\ref{xhat2}) in (\ref{xhat}) and using (\ref{Gauss_suff}) yields (\ref{T}). Since $T_j$ is a sufficient statistic of ${{\bf{y}}_j},{{{\bf{\hat x}}}_i},{{{\bf{\hat x}}}_l},...$ and ${{\bf{y}}_j},{{{\bf{\hat x}}}_i},{{{\bf{\hat x}}}_l},... \leftrightarrow {\bf{x}} \leftrightarrow {{\bf{y}}_k}$, it follows from Lemma \ref{Gauss1} that it is also a conditional sufficient statistic given ${{\bf{y}}_k}$. From Lemma \ref{lem3}, one can then replace ${{\bf{y}}_j},{{{\bf{\hat x}}}_i},{{{\bf{\hat x}}}_l},...$ by $T_j$ in (\ref{T}) and get the same RD function.

\end{IEEEproof}

At this point, we have shown that the above problem with multiple sources can be converted into a single source DSC problem for Gaussian sources with covariance distortion constraint. This problem is illustrated in Fig.\ref{prob}. For the case of mean-squared error distortion constraint, the RD function was found in \cite{Tian}, while the case of covariance distortion was not treated in that work. In the next section, we derive the RD function for the covariance distortion constraint under some mild technical assumptions.


\section{Rate-Distortion Function}

\begin{figure}[!t]
\centering
\includegraphics[scale=0.3]{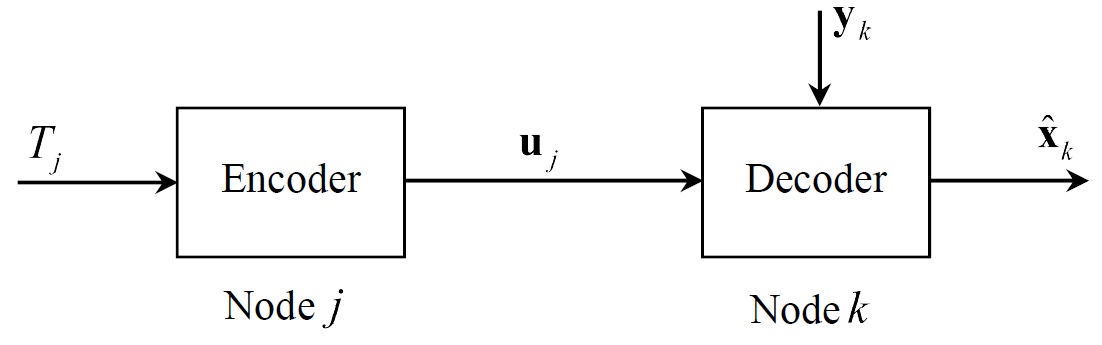}
\caption{The simplified DSC problem}
\label{prob}
\end{figure}

For the sake of simplicity of derivations, we write $\bf{x}$ and $T_j$ in terms of their linear estimations based on the known Gaussian vectors in Fig.\ref{prob}. In particular, we have that

\begin{eqnarray}
\label{tyu}
&&{T_j} = {\bf{A}}{{\bf{y}}_k} + {\bf{B}}{{\bf{u}}_j} + {{\bf{\upsilon }}_1},\\
\label{xty}
&&{\bf{x}} = {\bf{C}}{T_j} + {\bf{G}}{{\bf{y}}_k} + {{\bf{\upsilon }}_2} = \left( {{\bf{G}} + {\bf{CA}}} \right){{\bf{y}}_k} + {\bf{CB}}{{\bf{u}}_j} + {\bf{C}}{{\bf{\upsilon }}_1} + {{\bf{\upsilon }}_2},\\
\label{ty}
&&{T_j} = {\bf{\Gamma }}{{\bf{y}}_k} + {{\bf{\upsilon }}_3},
\end{eqnarray}

\noindent
where ${{\bf{\upsilon }}_1},{{\bf{\upsilon }}_2},{{\bf{\upsilon }}_3}$ are estimation errors with covariance matrices ${{\bf{\Sigma }}_{{T_j}|{{\bf{y}}_k},{{\bf{u}}_j}}},{{\bf{\Sigma }}_{{\bf{x}}|{T_j},{{\bf{y}}_k}}},{{\bf{\Sigma }}_{{T_j}|{{\bf{y}}_k}}}$, respectively, and ${\bf{A}},{\bf{B}},{\bf{C}},{\bf{G}}$ and ${\bf{\Gamma}}$ depend only on the covariance and cross-covariance matrices of ${\bf{x}}, {T_j}, {{\bf{y}}_k}$ and ${{\bf{u}}_j}$. (See Appendix for the mathematical statement for ${\bf{C}}$ as an example.) We will show that if the mixing matrices ${\bf{A}}_j$ in (\ref{measure}) are invertible\footnote{The mixing matrices are actually non-square and thus non-invertible. In this work we consider a square approximation. The general non-square case will be considered in our future work.}, then for ${{\bf{\Sigma }}_{x|{T_j},{{\bf{y}}_k}}} \prec {{\bf{D}}_j} \preceq {{\bf{\Sigma }}_{x|{{\bf{y}}_k}}}$ the RD function for node $j$ is given by:

\begin{equation}
\label{RD}
{R_j}\left( {{{\bf{D}}_j}} \right) = \frac{1}{2}\log \left( {\frac{{\left| {{{\bf{\Sigma }}_{x|{{\bf{y}}_k}}} - {{\bf{\Sigma }}_{x|{T_j},{{\bf{y}}_k}}}} \right|}}{{\left| {{{\bf{D}}_j} - {{\bf{\Sigma }}_{x|{T_j},{{\bf{y}}_k}}}} \right|}}} \right).
\end{equation}

\vspace{2mm}
\subsection{Lower Bound}
 First we need the following lemma:

\begin{lemma}\label{inv}
If the mixing matrices ${\bf{A}}_j, j=1,2,...$, are invertible, the matrix $\bf{C}$ in (\ref{xty}) is also invertible.
\end{lemma}

\begin{IEEEproof}
See the appendix.
\end{IEEEproof}

From the first equality in (\ref{xty}) we have:

\begin{equation}
\label{lb1}
{\bf{C}}{{\bf{\Sigma }}_{{T_j}|{{\bf{y}}_k}}}{{\bf{C}}^T} = {{\bf{\Sigma }}_{x|{{\bf{y}}_k}}} - {{\bf{\Sigma }}_{x|{T_j},{{\bf{y}}_k}}}.
\end{equation}

\noindent
From the second equaltiy in (\ref{xty}) we can write:

\begin{equation}
\label{new}
{\bf{x}} - {{{\bf{\hat x}}}_k} = {\bf{C}}{{\bf{\upsilon }}_1} + {{\bf{\upsilon }}_2}.
\end{equation}

\noindent
Assume now that the sequence of independent vectors ${\bf{x}}_i$ generated by the source is encoded in block vectors ${{\bf{X}}}$ each containing $N$ vectors. From the distortion constraint and (\ref{new}) we can write:

\begin{equation}
{{\bf{D}}_j} \succeq \frac{1}{N}\sum\limits_{i = 1}^N {E\left[ {\left( {{\bf{x}}_i - {\bf{\hat x}}_k^i} \right){{\left( {{\bf{x}}_i - {\bf{\hat x}}_k^i} \right)}^T}} \right]}  = {\bf{C}}{{\bf{\Sigma }}_{{{\bf{\upsilon }}_1}}}{{\bf{C}}^T} + {{\bf{\Sigma }}_{{{\bf{\upsilon }}_2}}}
\end{equation}

\noindent
or equivalently:

\begin{equation}
\label{lb2}
{\bf{C}}{{\bf{\Sigma }}_{{T_j}|{{\bf{u}}_j},{{\bf{y}}_k}}}{{\bf{C}}^T} \preceq {{\bf{D}}_j} - {{\bf{\Sigma }}_{x|{T_j},{{\bf{y}}_k}}}.
\end{equation}

\noindent
Denoting the blocks of $N$ vectors $T_j^i,{\bf{y}}_k^i,{\bf{u}}_j^i; \: i=1,...,N$ by ${{\bf{T}}_j},{{\bf{Y}}_k},{{\bf{U}}_j}$, respectively, we have:

\begin{eqnarray}
N{R_j}\left( {{D_j}} \right) &&\ge I\left( {{{\bf{T}}_j};{{\bf{U}}_j}|{{\bf{Y}}_k}} \right) \nonumber \\
 &&= h\left( {{{\bf{T}}_j}|{{\bf{Y}}_k}} \right) - h\left( {{{\bf{T}}_j}|{{\bf{U}}_j},{{\bf{Y}}_k}} \right) \nonumber \\
 &&= \sum\limits_{i = 1}^N {h\left( {T_j^i|{\bf{y}}_k^i} \right) - h\left( {T_j^i|{\bf{T}}_j^{i - 1},{\bf{u}}_j^i,{\bf{y}}_k^i} \right)} \nonumber \\
\label{lb5}
 &&\ge \sum\limits_{i = 1}^N {h\left( {T_j^i|{\bf{y}}_k^i} \right) - h\left( {T_j^i|{\bf{u}}_j^i,{\bf{y}}_k^i} \right)} \\
 &&= \frac{N}{2}\log \left( {\frac{{\left| {{{\bf{\Sigma }}_{{T_j}|{{\bf{y}}_k}}}} \right|}}{{\left| {{{\bf{\Sigma }}_{{T_j}|{{\bf{u}}_j},{{\bf{y}}_k}}}} \right|}}} \right) \nonumber \\
\label{lb3}
 &&= \frac{N}{2}\log \left( {\frac{{\left| {{\bf{C}}{{\bf{\Sigma }}_{{T_j}|{{\bf{y}}_k}}}{{\bf{C}}^T}} \right|}}{{\left| {{\bf{C}}{{\bf{\Sigma }}_{{T_j}|{{\bf{u}}_j},{{\bf{y}}_k}}}{{\bf{C}}^T}} \right|}}} \right)\\
\label{lb4}
 &&\ge \frac{N}{2}\log \left( {\frac{{\left| {{{\bf{\Sigma }}_{x|{{\bf{y}}_k}}} - {{\bf{\Sigma }}_{x|{T_j},{{\bf{y}}_k}}}} \right|}}{{\left| {{{\bf{D}}_j} - {{\bf{\Sigma }}_{x|{T_j},{{\bf{y}}_k}}}} \right|}}} \right),
\end{eqnarray}

\noindent
where the block vector ${{\bf{T}}_j^{i - 1}}$ contains $i-1$ vectors $T_j^1,...,T_j^{i - 1}$, (\ref{lb5}) is because conditioning reduces the entropy, (\ref{lb3}) is because $\bf{C}$ is invertible, and (\ref{lb4}) is the result of applying (\ref{lb1}) and (\ref{lb2}) to (\ref{lb3}).


\vspace{2mm}
\subsection{Upper Bound}

The lower bound derived in the previous part can be used as a guideline for the best possible performance for any coding scheme. The question is then, given the source $T_j$, how to encode it into a discrete source ${{\bf{u}}_j}$, so that for the given distortion constraint, the required rate for transmission of ${{\bf{u}}_j}$ achieves the lower bound?

Let us assume that $T_j$ is quantized to ${{\bf{u}}_j}$, in a way that satisfies the distortion constraint. From the results of DSC, due to the availability of the side information ${\bf{y}_k}$ at the decoder, it is possible to noiselessly encode ${{\bf{u}}_j}$ in blocks of sufficiently large length $N$ with a rate arbitrarily close to $H({{\bf{u}}_j}|{{\bf{y}}_k})=I\left( {{T_j};{{\bf{u}}_j}|{{\bf{y}}_k}} \right)$ \cite{SW}, \cite{Wyner1}, \cite{Wyner2}. The remaining task is to design a scheme for which $I\left( {{T_j};{{\bf{u}}_j}|{{\bf{y}}_k}} \right)$ achieves (\ref{RD}). This is possible by using the following scheme:

\begin{equation}
\label{scheme}
{{\bf{u}}_j} = {\bf{UC}}{T_j} + {{\bf{\nu }}_j},
\end{equation}

\noindent
where we have denoted the eigenvalue decomposition of $\left({{{\bf{\Sigma }}_{x|{{\bf{y}}_k}}} - {{\bf{\Sigma }}_{x|{T_j},{{\bf{y}}_k}}}}\right)$ by ${{\bf{U}}^{T}}{\bf{\Lambda}}{\bf{U}}$, and the covariance ${{\bf{\Sigma }}_{{{\bf{\nu }}_j}}}$ of the coding noise ${{\bf{\nu }}_j}$ is defined as \footnote{Note that it is a symmetric positive definite matrix.}:

\begin{equation}
\label{scheme1}
{{\bf{\Sigma }}_{{{\bf{\nu }}_j}}} = {\bf{U}}{\left({{{\bf{\Sigma }}_{x|{{\bf{y}}_k}}} - {{\bf{\Sigma }}_{x|{T_j},{{\bf{y}}_k}}}}\right)}{{\left( {{\bf{\Sigma }}_{x|{{\bf{y}}_k}}}-{\bf{D}}_{j} \right)}^{-1}}{\left({{{\bf{D}}_{j}} - {{\bf{\Sigma }}_{x|{T_j},{{\bf{y}}_k}}}}\right)}{\bf{U}^\textit{T}} ,
\end{equation}



\noindent
To verify this, one can write $I\left( {{T_j};{{\bf{u}}_j}|{{\bf{y}}_k}} \right)$ as $h\left( {{{\bf{u}}_j}|{{\bf{y}}_k}} \right) - h\left( {{{\bf{u}}_j}|{T_\textit{j}},{{\bf{y}}_k}} \right)$, substitute (\ref{scheme})--(\ref{scheme1}), and utilize (\ref{ty}) and (\ref{lb1}). 


It is worth noting that the RD function (\ref{RD}) generalizes the RD function of \cite{Wornell}, which treated the scalar case.


\vspace{5mm}
\section{Conclusions, Discussions, and Future Work}
We showed that the rate-distortion function for a distributed source coding problem with multiple sources at the encoder is identical to the rate-distortion function for the distributed encoding of a so-called conditional sufficient statistic of the sources. We derived a conditional sufficient statistic for the case that additive noises are Gaussian and mutually independent, and calculated the rate-distortion function in case that the sources are vector Gaussian and the distortion constraint is defined as a covariance matrix. Since vector sources were considered in order to take the memory into account, and the covariance constraint on the distortion is a more flexible fidelity criterion compared to mean-squared error, these results can be applied to the problem of source coding for audio signals in presence of reverberation, which will be the scope of our future work.


\appendix
Assume that the Gaussian vector $\bf{x}$ is estimated using two other Gaussian vectors $\bf{y}$ and $\bf{z}$ as:

\begin{equation}
{\bf{x = Cy + Gz + n}},
\end{equation}

\noindent
where the covariance matrix of the estimation error $\bf{n}$ is ${{\bf{\Sigma }}_n} = {{\bf{\Sigma }}_{x|yz}}$, and $\bf{y}$ and $\bf{z}$ are related to $\bf{x}$ via ${\bf{y}} = {\bf{Ax}+{\bf{n}}_1}$ and ${\bf{z}} = {\bf{Bx}+{\bf{n}}_2}$, and $\bf{A}$ and $\bf{B}$ are invertible. Therefore we have:

\begin{eqnarray}
\label{sig1}
&&{{\bf{\Sigma }}_{xy}} = {{\bf{\Sigma }}_x}{{\bf{A}}^T},\\
\label{sig2}
&&{{\bf{\Sigma }}_y} = {\bf{A}}{{\bf{\Sigma }}_x}{{\bf{A}}^T} + {{\bf{\Sigma }}_{{n_1}}},\\
\label{sig3}
&&{{\bf{\Sigma }}_{yz}} = {\bf{A}}{{\bf{\Sigma }}_x}{{\bf{B}}^T},\\
\label{sig4}
&&{{\bf{\Sigma }}_{xz}} = {{\bf{\Sigma }}_x}{{\bf{B}}^T}.
\end{eqnarray}

\noindent
From results in linear estimation theory, we obtain the following expression for $\bf{C}$ \cite{Est}:

\begin{eqnarray}
\label{ap1}
&&{\bf{C}}{{\bf{\Sigma }}_y} = {{\bf{\Sigma }}_{xy}} + \left( {{{\bf{\Sigma }}_{xy}}{\bf{\Sigma }}_y^{ - 1}{{\bf{\Sigma }}_{yz}} - {{\bf{\Sigma }}_{xz}}} \right){{\bf{\Delta }}^{ - 1}}{\bf{\Sigma }}_{yz}^T, \\
\label{ap2}
&&{\bf{\Delta }} = {{\bf{\Sigma }}_z} - {\bf{\Sigma }}_{yz}^T{\bf{\Sigma }}_y^{ - 1}{{\bf{\Sigma }}_{yz}}.
\end{eqnarray}

\noindent
The goal is to prove that ${\bf{C}}$ or equivalently ${\bf{C}}{{\bf{\Sigma }}_y}$ is invertible. Substituting (\ref{sig1})--(\ref{sig4}) in (\ref{ap1}) yields:

\begin{eqnarray}
{\bf{C}}{{\bf{\Sigma }}_y} &&= {{\bf{\Sigma }}_x}{{\bf{A}}^T} + \left( {{{\bf{\Sigma }}_x}{{\bf{A}}^T}{\bf{\Sigma }}_y^{ - 1}{\bf{A}}{{\bf{\Sigma }}_x}{{\bf{B}}^T} - {{\bf{\Sigma }}_x}{{\bf{B}}^T}} \right){{\bf{\Delta }}^{ - 1}}{\bf{\Sigma }}_{yz}^T \nonumber \\
 &&= {{\bf{\Sigma }}_x}{{\bf{A}}^T} + {{\bf{\Sigma }}_x}\left( {{{\bf{A}}^T}{\bf{\Sigma }}_y^{ - 1}{\bf{A}}{{\bf{\Sigma }}_x} - {\bf{I}}} \right){{\bf{B}}^T}{{\bf{\Delta }}^{ - 1}}{\bf{\Sigma }}_{yz}^T \nonumber \\
 &&= {{\bf{\Sigma }}_x}{{\bf{A}}^T} + {{\bf{\Sigma }}_x}{{\bf{A}}^T}\left( {{\bf{\Sigma }}_y^{ - 1} - {{\bf{A}}^{ - T}}{\bf{\Sigma }}_x^{ - 1}{{\bf{A}}^{ - 1}}} \right){\bf{A}}{{\bf{\Sigma }}_x}{{\bf{B}}^T}{{\bf{\Delta }}^{ - 1}}{\bf{\Sigma }}_{yz}^T \nonumber \\
 &&= {{\bf{\Sigma }}_x}{{\bf{A}}^T} + {{\bf{\Sigma }}_x}{{\bf{A}}^T}\left[ {{{\left( {{\bf{A}}{{\bf{\Sigma }}_x}{{\bf{A}}^T} + {{\bf{\Sigma }}_{{n_1}}}} \right)}^{ - 1}} - {{\left( {{\bf{A}}{{\bf{\Sigma }}_x}{{\bf{A}}^T}} \right)}^{ - 1}}} \right] 
 {{\bf{\Sigma }}_{yz}}{{\bf{\Delta }}^{ - 1}}{\bf{\Sigma }}_{yz}^T \nonumber \\
\label{ap3}
 &&= {{\bf{\Sigma }}_x}{{\bf{A}}^T}\left( {{\bf{I}} + {{\bf{H}}^{ - 1}}{{\bf{\Sigma }}_{yz}}{{\bf{\Delta }}^{ - 1}}{\bf{\Sigma }}_{yz}^T} \right),
\end{eqnarray}

\noindent
where $\bf{H}$ is defined as:

\begin{equation}
{{\bf{H}}^{ - 1}} = {\left( {{\bf{A}}{{\bf{\Sigma }}_x}{{\bf{A}}^T} + {{\bf{\Sigma }}_{{n_1}}}} \right)^{ - 1}} - {\left( {{\bf{A}}{{\bf{\Sigma }}_x}{{\bf{A}}^T}} \right)^{-1}}.
\end{equation}

\noindent
From (\ref{ap2}) and (\ref{sig1})--(\ref{sig4}) we have:

\begin{eqnarray}
{\bf{\Delta }} &&= {\bf{B}}{{\bf{\Sigma }}_x}{{\bf{B}}^T} + {{\bf{\Sigma }}_{{n_2}}} - {\bf{B}}{{\bf{\Sigma }}_x}{{\bf{A}}^T}{\left( {{\bf{A}}{{\bf{\Sigma }}_x}{{\bf{A}}^T} + {{\bf{\Sigma }}_{{n_1}}}} \right)^{ - 1}}{\bf{A}}{{\bf{\Sigma }}_x}{{\bf{B}}^T} \nonumber \\
 &&= {\bf{B}}{{\bf{\Sigma }}_x}\left[ {{\bf{I}} - {{\bf{A}}^T}{{\left( {{\bf{A}}{{\bf{\Sigma }}_x}{{\bf{A}}^T} + {{\bf{\Sigma }}_{{n_1}}}} \right)}^{ - 1}}{\bf{A}}{{\bf{\Sigma }}_x}} \right]{{\bf{B}}^T} + {{\bf{\Sigma }}_{{n_2}}} \nonumber \\
 &&= {\bf{B}}{{\bf{\Sigma }}_x}{{\bf{A}}^T}\left[ {{{\bf{A}}^{ - T}}{\bf{\Sigma }}_x^{ - 1}{{\bf{A}}^{ - 1}} - {{\left( {{\bf{A}}{{\bf{\Sigma }}_x}{{\bf{A}}^T} + {{\bf{\Sigma }}_{{n_1}}}} \right)}^{ - 1}}} \right]
 {\bf{A}}{{\bf{\Sigma }}_x}{{\bf{B}}^T} + {{\bf{\Sigma }}_{{n_2}}} \nonumber \\
\label{ap4}
 &&= {\bf{\Sigma }}_{yz}^T\left( { - {{\bf{H}}^{ - 1}}} \right){{\bf{\Sigma }}_{yz}} + {{\bf{\Sigma }}_{{n_2}}}.
\end{eqnarray}

\noindent
Substituting (\ref{ap4}) in (\ref{ap3}) and defining $\bf{P}$ as:

\begin{equation}
{{\bf{P}}^{ - 1}} = {\left( {{\bf{\Sigma }}_{yz}^T{{\bf{H}}^{ - 1}}{{\bf{\Sigma }}_{yz}}} \right)^{ - 1}} - {\left( {{\bf{\Sigma }}_{yz}^T{{\bf{H}}^{ - 1}}{{\bf{\Sigma }}_{yz}} - {{\bf{\Sigma }}_{{n_2}}}} \right)^{ - 1}},
\end{equation}

\noindent
we have:

\begin{equation}
{\bf{C}}{{\bf{\Sigma }}_y} = {{\bf{\Sigma }}_x}{{\bf{A}}^T}{{\bf{H}}^{ - 1}}{{\bf{\Sigma }}_{yz}}{{\bf{P}}^{ - 1}}{\bf{\Sigma }}_{yz}^T,
\end{equation}

\noindent
which is clearly invertible.


\end{document}